# The structural origin of the hard-sphere glass transition in granular packing

Chengjie Xia[1], Jindong Li[1], Yixin Cao[1], Binquan Kou[1], Xianghui Xiao[2], Kamel Fezzaa[2], Tiqiao Xiao[3] & Yujie Wang[1,4]*

[1]*Department of Physics and Astronomy, Shanghai Jiao Tong University, 800 Dong Chuan Road, Shanghai 200240, China*
[2]*Advanced Photon Source, Argonne National Laboratory, 9700 South Cass Avenue, Argonne, IL 60439, USA*
[3]*Shanghai Institute of Applied Physics, Chinese Academy of Sciences, Shanghai 201800, China*
[4]*Collaborative Innovation Center of Advanced Microstructures, Nanjing University, Nanjing, 210093, China*

**Abstract**:

Glass transition is accompanied by a rapid growth of the structural relaxation time and a concomitant decrease of configurational entropy. It remains unclear whether the transition has a thermodynamic origin, and whether the dynamic arrest is associated with the growth of a certain static order. Using granular packing as a model hard-sphere glass, we clearly identified the glass transition as a thermodynamic phase transition with a 'hidden' polytetrahedral order. This polytetrahedral order is spatially correlated with the slow dynamics. It is geometrically frustrated and has a peculiar fractal dimension. Additionally, as the packing fraction increases, its growth follows an entropy-driven nucleation process, similar to that of the random first-order transition theory. Our study essentially identifies a long-sought-after structural glass order in hard-sphere glasses.

When a liquid is cooled towards the glass transition, the dynamics slows down dramatically. The mechanism of this phenomenon has been extensively investigated for decades, but there is no consensus on it yet[1,2]. Recently, the discovery of dynamic heterogeneity and the rapid increase of its correlation length near the glass transition suggest the collective nature of the dynamics, which has inspired hope that a corresponding static correlation length associated with some critical behavior similar to an ordinary phase transition can be identified[2,3]. This length scale is also supposed to have a configurational entropy origin, as originally suggested by the Adam-Gibbs theory[4]. Recent searches for this static correlation length have identified crystalline, icosahedron-like, or 'point-to-set' type of orders which show significant increases of the correlation lengths as dynamic arrest is approached[5]. However, for all existing approaches, either the relationship with the configurational entropy or the structural nature of the order remains to be understood.

The hard-sphere system is a popular model glass former because it can simulate systems such as metallic glasses, granular systems, and colloidal suspensions[6]. The close relationship between static granular packing and hard-sphere glasses dates back to the pioneering work of Bernal, who experimentally investigated granular packing to model liquid structures[7]. Additionally, agitated granular systems exhibit slow dynamics that resemble those of thermal glassy systems[8-11]. The analogy between granular packing and a thermal glass can be formally understood by Edwards' ensemble, on which a statistical framework similar to the equilibrium statistical mechanics can be established for static granular packing[12]. Additionally, the hard-sphere system has a zero-temperature geometric phase transition, the jamming transition, which demonstrates interesting properties like marginal stability and critical scaling behaviors[13,14]. Jamming transition and its distinction with the glass transition have recently been characterized from both thermodynamic[15] and rheological[16] points of view. A recent theoretical approach has tried to incorporate the jamming transition into the framework of glass theory through a Gardner transition to fractal sub-basins in the free energy landscape[17]. However, the possible structural changes associated with these two transitions and the structural nature of the fractal sub-basins remain to be explored.

Using granular packing as a model hard-sphere glass system, we identified a geometrically frustrated polytetrahedral structural order which fulfills the requirements of a static glass order. By using synchrotron X-ray imaging techniques[18,19], we carried out a systematic study of the dynamics, thermodynamics, and structures of tapped granular packing. We demonstrate that the system exhibits all key phenomena of a thermal glassy system. Particularly, a polytetrahedral structural order grows rapidly as the packing fraction increases and it is spatially correlated with the slow relaxation dynamics. The non-trivial fractal dimension and length scale of this polytetrahedral order are consistent with an entropy-driven nucleation model similar to the random first-order transition (RFOT) theory[20].

## Results

**Thermodynamic variables based on Edwards' ensemble.** To obtain the thermodynamic variables and detailed local structures of the packing, we carry out X-ray micro-tomography scans on packing of different average packing fractions $\Phi$ prepared by three different protocols (Methods). Based on the statistical framework originally proposed by Edwards and coworkers[12], the thermodynamic variables, such as entropy $S$ and compactivity $\chi$ (similar role as temperature), are evaluated from the fluctuations of the reduced Voronoi cell volumes $v_{voro} = V_{voro}/V_g$, where $V_g$ and $V_{voro}$ are the particle volumes and their Voronoi cell volumes respectively, according to[21,22]

$$\frac{1}{\chi(\Phi)} - \frac{1}{\chi(\Phi=0.55)} = \frac{1}{V_g}\int_{0.55}^{\Phi}\frac{d\Phi'}{\Phi'^2 \sigma_v^2} \tag{1}$$

and

$$S(\Phi) - S(\Phi=0.64) = V_g\int_{\Phi}^{0.64}\frac{d\Phi'}{\Phi'^2 \chi(\Phi')}, \tag{2}$$

where $\sigma_v^2 = \langle v_{voro}^2\rangle - \langle v_{voro}\rangle^2$. The constant, which is similar to the Boltzmann constant, is set to unity. $V_g$ is also set to unity for convenience. We further assume that the compactivity of random loose packing is infinitely large, that is, $\frac{1}{\chi(\Phi=0.55)} = 0$, to complete the integral of equation (1). In equation (2), the entropy $S(\Phi)$ can only be identified up to a constant and we assume $S(\Phi=0.64) \approx 1.1$ by using the Shannon entropy calculation results[22]. The integral of equation (1) is calculated numerically starting from a second-order polynomial fitting of $\sigma_v^2$ as a function of $\Phi$: $\sigma_v^2 = 2.387\Phi^2 - 3.063\Phi + 0.997$ (Fig. 1a). Our results are in quantitative agreements with previous study[22] (Fig. 1b). It is worth noting that the configurational entropy defined above corresponds to the complexity in hard-sphere glass terminology and should not be confused with the definition as the entropy difference between liquid and crystal phases[6].

Like a thermal hard-sphere glass, $S$ decreases with decreasing $\chi$ or increasing $\Phi$. We extrapolate $S(\Phi)$ and $S(\chi)$ curves to $S=0$ to obtain the corresponding glass-close-packing (GCP) packing fraction $\Phi_{GCP}=0.671$ or GCP compactivity $\chi_{GCP}=0.0554$ (refs 6,23,24). $\Phi_{GCP}$ is close to the jammed ideal glass transition density 0.68 from replica theory calculations[6]. In the following, structural relaxation time and structural correlation length will be expressed as functions of these thermodynamic variables to draw analogies between the tapped granular system and a thermal hard-sphere glass.

**Relaxation time.** To study the slow relaxation dynamics in granular packing, we use X-ray absorption imaging to measure the time evolution of the average packing fraction $\Phi$ under tapping (Methods). The structural relaxation times $\tau$ are calculated from the compaction curves at different tapping intensities $\Gamma$. The packing is first tapped at $\Gamma = 15$ for 1,000 times to reach $\Phi_0 = 0.615$. Then, the compaction curves are measured by tapping the packing at different $\Gamma$ to reach the corresponding reversible-branch packing fractions $\Phi_\infty = \Phi(\Gamma)$ (Fig. 1c). The packing that has reached reversible-branch is at steady state and memoryless. Each compaction curve can then be fitted using the empirical Kohlrausch-Williams-Watts law[8]: $\Phi(t) = \Phi_\infty - (\Phi_\infty - \Phi_0)\exp(-(t/\tau)^\beta)$ to obtain $\tau$, where $t$ is the number of taps and $\beta$ is the stretching exponent (Fig. 1d). The fitted values of $\beta$ lie in the range of 0.5-0.9 if both $\tau$ and $\beta$ are allowed to vary. To be consistent, we fix $\beta = 0.7$ to obtain $\tau$ for different compaction curves. Nonetheless, the $\tau$ values remain essentially unchanged as compared with the case when $\beta$ is allowed to vary. As shown in Fig. 2, $\tau$ increases with decreasing $\chi$ or increasing $\Phi$, similar to a thermal hard-sphere glass[25].

**Polytetrahedral order.** As shown above, the great analogies in slow dynamics and thermodynamics between tapped granular packing and a thermal hard-sphere glass suggest a common origin. In the following, we propose that the formation of local geometrically frustrated quasi-regular tetrahedra is the microscopic mechanism for the dynamic arrest in both systems[26,27]. We demonstrate that the polytetrahedral order associated with these quasi-regular tetrahedra corresponds to the long-sought-after glass order in hard-sphere glasses[26,27], by showing: the polytetrahedral order is spatially correlated with the slow relaxation dynamics; the static correlation length of this polytetrahedral order increases rapidly as $\Phi$ increases; and the size and shape of the polytetrahedral order are consistent with a configurational entropy-driven nucleation model similar to the RFOT theory[20].

Similar to previous studies[28], a quasi-regular tetrahedron is defined as a Delaunay simplex whose shape is close to a regular tetrahedron, with the shape deviation less than some threshold value of a polytetrahedral order parameter $\delta = e_{max} - 1$. In this expression, $e_{max}$, in units of mean particle diameter $\sigma$, is the length of the longest edge of the tetrahedron. The $\delta$ values lie in a range between zero and an upper limit around one. There exist other measures to define a quasi-regular tetrahedron, such as tetrahedricity $\Delta$ (ref. 23). However, it turns out that our general results are not sensitive to any particular definition.

**Correlation between order and dynamics.** We first show that the quasi-regular tetrahedra are spatially correlated with the slow relaxation dynamics. In the tapping experiment with $\Gamma = 4$, we track the trajectories of all particles for a number of taps by conducting a tomographic scan after each tap (Methods). We define $\Delta \mathbf{r}_i = d\mathbf{r}_i - \mathbf{v}(\mathbf{r}_i)$ as the diffusing displacement after one tap of particle $i$ located at $\mathbf{r}_i$, where the absolute displacement $d\mathbf{r}_i$ is subtracted by an averaged steady-state convection displacement $\mathbf{v}(\mathbf{r}_i)$ (Methods). To characterize the mobility of particles belonging to tetrahedra of different $\delta$ values, we define the $\delta$-mobility $\Delta r^2(\delta) = \left\langle |\Delta \mathbf{r}_i|^2 \right\rangle_\delta$ (ref. 29), where the average is taken over all particles composing tetrahedra with $\delta$. As shown in Fig. 3a, a positive correlation between $\delta$-mobility and $\delta$ value is clearly visible. The average value of $\sqrt{\Delta r^2}$ after one tap is about $0.04\sigma$, which is close to the typical cage size in the packing[30]. This is owing to the fact that one tap duration at $\Gamma = 4$ is on similar timescale as the structural relaxation time $\tau$, that is, the particles have undergone many collisions upon one tap. It's worth noting that similar correlation between the particle mobility and the shape of tetrahedron it belongs to has also been observed in colloidal systems[31], which suggests that the inherent friction of granular system is not the cause of the structure-dynamics correlation.

As an alternative evidence, we define a tetrahedron correlation function $p_\delta(t)$ as the probability that one tetrahedron (Delaunay simplex) with $\delta$ at $t = 0$ is composed of the same four particles from $t = 0$ to the $t^{\text{th}}$ tapping. This function captures the dependency of the local structural relaxation time upon $\delta$. As shown in Fig. 3b, tetrahedra with smaller $\delta$ relax much more slowly than those with larger $\delta$.

**Spatial correlation of polytetrahedral order.** The fraction of quasi-regular tetrahedra grows as $\Phi$ increases, which is accompanied by increasing spatial correlations among them, that is, they tend to aggregate with each other. This spatial correlation can be demonstrated explicitly in terms of a percolation analysis on the Delaunay networks[32]. Tetrahedra are colored according to their $\delta$ values (tetrahedra with $\delta < \delta_c$ are colored where $\delta_c$ is a threshold) or colored randomly but with the same number of tetrahedra. In both cases, face-adjacent colored tetrahedra are joined together to form clusters. We cut a cubic region out of the packing and define $L_{\text{cluster}}$ as the longest spanning range of each cluster in directions parallel to the three axes of the cube. $\text{Max}(L_{\text{cluster}})$ grows as more cells can be colored when the threshold value of $\delta_c$ is gradually relaxed, and it can ultimately reach the size of the cubic region $L_{\text{box}}$ at the percolation limit. As shown in Fig. 4, it turns out that tetrahedra chosen based on their $\delta$ (or $\Delta$) values show higher likelihood to percolate compared with the random-coloring case, which suggests their spatial correlations. To emphasize the unique relevancy of this polytetrahedral order, similar percolation analyses are carried out over various other structural order parameters, which show almost identical percolation behaviors with their random counterparts, suggesting that there exist negligible correlations[29] (Methods). Next, we use $\delta_c = \delta^* = 0.245$ to select out quasi-regular tetrahedra at different $\Phi$. The threshold value adopted is similar to previous studies[28,33]. The polytetrahedral order associated with these quasi-regular tetrahedra has a polytetrahedral structure[28]. The specific $\delta^*$ value chosen corresponds to a percolation transition of the polytetrahedral order at random close packing (RCP) ($\Phi = 0.64$) associated with the jamming transition of frictionless particles[13,28,34].

We quantify how the spatial correlations among these quasi-regular tetrahedra vary with $\Phi$ by calculating the correlation length $\xi$ at different $\Phi$. The average size of the polytetrahedral clusters $\xi_c$ is evaluated using the radius of gyration $R_g$ of all clusters, $\xi_c^2 = \dfrac{2 \sum R_g^2 N_c^2}{\sum N_c^2}$, where $N_c$ is the number of particles belonging to a cluster, and $R_g^2 = \dfrac{1}{2} \langle |\mathbf{r}_i - \mathbf{r}_j|^2 \rangle_{i,j}$ is half the average square distance between all pairs of particles in a cluster[35]. We note that $\xi_c$ includes contributions from both the intrinsic correlation length $\xi$ of the polytetrahedral order and a trivial correlation length $\xi_r$ associated with the random percolation process, which also increases with $\Phi$ (inset of Fig. 5a). Therefore we define $\xi = \xi_c - \xi_r$ which equals to zero when the tetrahedra are uncorrelated. As shown in Fig. 5a, $\xi$, in units of $\sigma$, also increases with decreasing $\chi$ or increasing $\Phi$, similar to $\tau$.

To further characterize the structural change associated with the growth of $\xi$, we study the evolution of $P$, which is the fraction of particles belonging to at least one quasi-regular tetrahedron. $P$ increases from 80% to 98% as $\Phi$ increases from 0.572 to 0.634 (Fig. 5b), whereas $\xi$ increases from 0.44 to 8.14 accordingly. This almost twenty-fold growth of $\xi$ can therefore only be induced by the merging of smaller polytetrahedral clusters into bigger ones instead of the simple inclusion of more particles. Meanwhile, the fraction of quasi-regular tetrahedra increases from 13% to 27%, as shown in Fig. 6a-d.

**Dependencies of $\tau$ and $\xi$ on thermodynamic variables.** In the following, we analyze the dependencies of $\tau$, $\xi$ on thermodynamic variables $\chi$ and $S$, similar to those in a thermal glassy system.

The relation between $\tau$ and $\chi$ can be well described by the Vogel-Fulcher-Tammann form (Fig. 2)

$$\tau = \tau_0 \exp\left(\frac{D\chi_\tau}{\chi - \chi_\tau}\right). \qquad (3)$$

The fitted $\chi_\tau = 0.049 \pm 0.003$ agrees nicely with the 0.045 value from previous hard-sphere simulation[25]. The corresponding $\Phi_\tau = 0.678$ is consistent with $\Phi_{GCP}$ from above entropy extrapolation[1,6], suggesting a similar relationship between dynamics and thermodynamics as a thermal hard-sphere glass[1]. $\tau_0 = 0.024 \pm 0.014$ tap turns out to be the microscopic timescale in our system which is much smaller than either the tap duration or the structural relaxation time $\tau$.

Additionally, $\tau$ and $S$ follows an Adam-Gibbs type of relation

$$\ln\frac{\tau}{\tau_0} \propto \left(\frac{1}{\chi s_c}\right)^{\frac{\theta\psi}{d-\theta}}, \qquad (4)$$

where $s_c$ is the configurational entropy density, defined as $S/\langle V_{voro}\rangle$, with $\langle V_{voro}\rangle$ being the average Voronoi cell volume (Fig. 2). The exponent $\frac{\theta\psi}{d-\theta} = 1.0 \pm 0.2$ is surprisingly close to the Adam-Gibbs relation[4]. Here, the exponent $\frac{\theta\psi}{d-\theta}$ follows the convention of the RFOT theory[20,36].

$\xi$ also shows a diverging behavior with decreasing $\chi$ that can be fitted using a power-law function[37]

$$\xi \propto \left(\frac{\chi_\xi}{\chi - \chi_\xi}\right)^\nu, \qquad (5)$$

which yields $\chi_\xi = 0.051 \pm 0.016$ ($\Phi_\xi = 0.675 \pm 0.023$) and $\nu = 1.4 \pm 0.6$ (Fig. 5a). The value of $\nu$ is different from the critical exponent ($\approx 2/3$) of the 3D Ising universality class as suggested in recent experiments[38].

We further demonstrate that the growth of the polytetrahedral order is consistent with a configurational entropy-driven nucleation model similar to the RFOT theory[20]. In RFOT theory, the static correlation length or the mosaic size $\xi_{mosaic}$, is determined by a competition between the gain in configurational entropy $S$ and a free energy cost proportional to $\xi_{mosaic}^\theta$ due to surface mismatch between different mosaics[20]. Specifically, the competition results in a typical length scale $\xi_{\text{mosaic}} \propto \left(\frac{1}{Ts_c}\right)^{\frac{1}{d-\theta}}$, where $T$ is the temperature and $d$ is the spatial dimension[37]. Motivated by this nucleation model, we plot $\xi$ versus $\frac{1}{\chi s_c}$ and fit the data according to

$$\xi \propto \left(\frac{1}{\chi s_c}\right)^{\frac{1}{d-\theta}} \tag{6}$$

(Fig. 5c). The fitting yields $\frac{1}{d-\theta} = 1.78 \pm 0.34$ ($\theta = 2.44 \pm 0.11$). At first sight, the $\theta$ value obtained was incompatible with the original RFOT theory ($\theta = d/2$)[37] or Adam-Gibbs relation ($\theta = 0$)[4]. However, as shown in Fig. 6e, this discrepancy can be naturally reconciled with the fractal nature of our polytetrahedral order[39], since it has a fractal surface dimension $\theta_s = 2.57$ (for $\Phi = 0.634$ packing) which is compatible with the extracted $\theta$ from above nucleation model (Fig. 6e). Notably, this unusual $\theta$ value has been observed before in both experiments and simulations[36]. Interestingly, $\theta_s$ decreases slightly with increasing $\Phi$, suggesting that the polytetrahedral order will have less rough surfaces as $\Phi$ increases (Fig. 6f). These fractal polytetrahedral clusters fail to tile space because they are frustrated geometrically[27,40]. Furthermore, cluster size $N_c$ shows a power-law distribution: $p(N_c) \propto N_c^{-\mu}$. It turns out that the hyperscaling relationship $d_f(\mu - 1) = d$ roughly holds, where $d_f$ is the cluster fractal dimension (Fig.6e). Similar fractal clusters have been observed for immobile particle clusters in other glass-forming liquids[41].

In addition to the proceeding cluster-analysis, we also attempt to extract the correlation length using a spatial correlation function of $\delta$ (Fig. 5d). This function behaves rather similarly to the standard pair correlation function $g(r)$ (ref. 19), which displays a finite-length decaying behavior but no discernable differences for packing with different $\Phi$.

This indicates that protocols based on pair correlation analysis are incapable of capturing correlations in our system[34]. Same problem could also exist in the 'point-to-set' type of analysis, because pinning a smooth boundary might not be the best way to capture a fractal phase inside[42].

## Discussion

The main conclusion of the current study is that quasi-regular tetrahedra are the structural elements of glass order in weakly-polydisperse hard-sphere glass-particle systems. The order grows by following an entropy-driven nucleation process which is reminiscent of the diffusion limited cluster aggregation (DLCA) in kinetic gelation process, where independent clusters following DLCA growth processes touch in forming a global percolating fractal structure and acquire mechanical rigidity suddenly[43]. The fact that the growth is correlated can therefore induce cooperativity and non-Arrhenius behavior in the system.

Similar to the gelation process in systems with attractive interactions[18,43], we suggest that the jamming transition corresponds to a rigidity percolation transition of glass order for systems with repulsive interactions, that is, at RCP, the percolated polytetrahedral clusters acquire an infinite mechanical correlation length abruptly. The fractal shape of the percolated polytetrahedral order could therefore be related to the marginal solid behavior[44] and unique scaling behaviors of the jamming transition[13]. Our jamming transition picture therefore suggests a unified scenario for rigidity transitions in systems with attractive or purely repulsive interactions: both are driven by the rigidity percolation of an underlying glass order.

The above percolation mechanism of jamming transition can also provide a simple geometric explanation of the continuous range of jamming density (the J-line) observed in numerical simulations[6,15,45]. In our system, the highest packing density we can theoretically achieve is approximately 0.64, which corresponds to the RCP state. Interestingly, the RCP state has a non-zero entropy and consists of many polytetrahedral clusters, and therefore is not the ideal glass state[46]. At RCP, the average internal packing fraction of each cluster is approximately 0.67, which suggests that the rather low global packing fraction originates from the existence of cluster boundaries. As a result, if we can extrapolate the configurational entropy towards zero in obtaining a single large polytetrahedron spanning the whole system, that is, the jammed ideal glass state or GCP, then in principle, we can obtain a jammed packing density around 0.67. Interestingly, this is exactly the upper limit of the J-line as has been predicted by the simulations[6,45].

Additionally, since the packing in the current work is close to the lower-density-limit of the J-line, their equilibrium counterparts correspond to supercooled liquid states which are not very deep in the free energy landscape. This supercooled liquid picture is consistent with both the mean-field study by Mari *et. al.* in which they found that the J-point correspond to the system just entering the landscape regime[45], and the fractal cooperatively rearranging regions (CRRs) found in glass-forming liquids near the dynamical crossover temperature[39], where the CRRs bear great similarity to the fractal polytetrahedra in our system. The location in the free energy landscape also naturally explains our anomalous scaling exponent $\theta$ as compared to that of the original RFOT theory, which mainly deals with mosaic states very deep in the free energy landscape.

In the current study, we implicitly assume the validity of Edwards' ensemble for packing prepared by both tapping and flow pulse protocols[21,47]. Despite the fact that packing prepared under these protocols has previously been established as ergodic, history-independent, and is therefore prone to a valid statistical analysis[21,47], there still exist some ongoing debates regarding the basic assumptions of the Edwards' framework, especially the flat measure assumption[48-53]. As of today, a direct correspondence between the Edwards' entropy and the configurational entropy of a thermal hard-sphere glass is still waiting to be established theoretically. Therefore, despite the self-consistency of a thermodynamic analysis of our granular hard-sphere glass and its great analogy with a thermal hard-sphere glass, a direct one-to-one correspondence should not simply be taken for granted. This should also be born in mind when comparing our entropy-driven nucleation picture with RFOT theory.

Overall, our current study illustrates the origin of fragile glass behavior in one type of model glass former. It suggests that other fragile glass systems can potentially be categorized by the growth of different types of structural orders, similar to the studies of crystalline orders.

## METHODS

**X-ray projection imaging.** We use X-ray projection absorption imaging to measure the average packing fraction $\Phi$. The granular particles used for all experiments in this study are glass particles (Duke Scientific, USA) with $200 \pm 15\,\mu m$ particle diameter and a slight polydispersity of around 3%. The experiment is conducted with a 27-keV monochromatic X-ray beam at the BL13W1 beamline of the Shanghai Synchrotron Radiation Facility (SSRF). $\Phi$ are measured at ten different tapping numbers (evenly spaced on the logarithmic scale) for each compaction curve.

To obtain $\Phi$, we also take flat-field images when the packing is outside the X-ray field-of-view. The projection images with and without the packing have intensity distributions $I(x,z)$ and $I_0(x,z)$, respectively, where $(x,z)$ denotes the coordinates of a pixel. According to Beer's light absorption law, $\Phi$ of the packing can be calculated as

$$\Phi = \frac{l_0 \iint \ln \frac{I_0(x,z)}{I(x,z)} dxdz}{\pi R^2 H}, \tag{7}$$

where $l_0$ is the attenuation length of glass at an X-ray energy of 27 keV, $R$ and $H$ are the radius and height of the container in the field-of-view. The integration is taken over the area covered by glass particles. The influence of the acrylic container has also been corrected. The $\Phi$ values obtained are consistent with independent tomography measurements as shown in Fig. 1c.

**X-ray micro-tomography.** We use micro-tomography to obtain the 3D packing structures prepared by tapping, hopper deposition, and flow pulse protocols. These three protocols can cover a wide range of $\Phi$ from 0.572 to 0.634. Using micro-tomography, we also investigate the correlation between dynamics and structure by tracking the displacements of all particles and the corresponding structural evolution of the packing for a consecutive number of tapping steps in the tapping experiment.

In the tapping protocol, we fill a 9-mm ID acrylic cylindrical container with particles to approximately 1 cm in height, and we use an electromagnetic exciter to tap the container. Packing with different $\Phi$ ranging from 0.618 to 0.634 is obtained by varying the tapping intensity $\Gamma$, which is measured by an accelerometer as the ratio between the peak-to-peak acceleration and the gravitational acceleration. The tapping consists of a single cycle of 60-Hz sine wave spaced with 0.5 s intervals to allow the system to relax completely. A total of 1000 taps are applied on each packing with different $\Gamma$ to reach steady state. In the hopper deposition protocol, we first place a hopper with its outlet touching the base of the cylindrical container, and fill the hopper with particles. The hopper is then slowly lifted up with a step motor to let the particles drain gradually from the outlet. The packing formed is cylindrical at the bottom, with a conical top. The $\Phi$ prepared by this protocol is 0.598. In the flow pulse protocol, the particles are placed in an acrylic cylindrical tube filled with water. The tube is sealed with a fine copper mesh at the bottom from a water inlet. The packing is prepared by subjecting the particles to a sequence of flow pulses generated from a syringe pump. The particles are allowed to fully settle between pulses. $\Phi$ ranging from 0.572 to 0.588 are obtained by varying the flow velocity.

The X-ray micro-tomography experiment is carried out at both the 2BM beamline of the Advanced Photon Source (APS) at Argonne National Laboratory, and the BL13W1 beamline of SSRF. At APS, the 'pink' X-ray beam from a bending magnet source with a median energy of 27 keV is used for the high-speed tomography image acquisitions, and the single exposure time is 3 ms. Each tomographic scan consists of 1500 projection images. At SSRF, the monochromatic 27-keV X-ray beam is used and the single exposure time is 40 ms. Each tomographic scan consists of 720 projection images. The 3D structures are first reconstructed using the conventional filtered back-projection algorithm. A marker-based watershed image segmentation technique is then implemented to obtain all particles' positions and sizes.

For the packing prepared using the tapping and flow pulse protocols, each reconstructed 3D structure consists of approximately 17,000 particles after excluding particles within four particle diameters from the container boundary. For the packing prepared using the hopper deposition protocol, approximately 2,600 particles are used after excluding those in the conical top and boundaries. For each reconstructed packing structure, we conduct a Voronoi tessellation, and the average packing fraction $\Phi$ is obtained by averaging the local packing fractions $\Phi_{loc}$, which is defined as the ratio between the volume of each particle and its Voronoi cell.

**Calculation of the convection displacement $\mathbf{v}(\mathbf{r}_i)$.** We coarse-grain the whole packing into sub-volumes of cubic shape with the size of each cube about three particle diameters. Since the convection is steady after extensive tapping, we calculate the convection displacement by simply averaging spatially and temporarily of the displacements of all particles inside each cube during the full tapping sequence. We also prove that the results are not sensitive to the coarse-graining size by varying the cube size from two to five particle diameters, and it turns out that the value of $\Delta r^2$ only slightly depends on the cube size, and its correlation with $\delta$ remains approximately unchanged.

**Percolation of various structural orders.** Various structural mechanisms of the glass transition have suggested the existence of different local structural orders, such as icosahedral order, crystalline order as the driving mechanism of the glass transition. To test all these different mechanisms, we define their corresponding structural order parameters to determine whether there exist significant increases in their spatial correlation lengths as $\Phi$ increases. The spatial correlations are analyzed based on percolation analyses on the Delaunay or Voronoi networks of the packing. In both networks, each cell is identified as a site, and the common surface between two sites as a bond. Sites are colored according to the values of various structural order parameters, or they are colored randomly. Colored sites that are connected to each other through bonds can form clusters, and they finally percolate the whole network when enough sites are colored.

We define two structural order parameters, $\delta$ and tetrahedricity $\varDelta$ based on the Delaunay cells[23].

$\delta = e_{max} - 1$, where $e_{max}$ is the longest tetrahedron edge in units of average particle diameter.

Tetrahedricity $\varDelta = \dfrac{\sum_{i<j}(e_i - e_j)^2}{15\langle e \rangle^2}$, where $\langle e \rangle = \dfrac{\sum_i e_i}{6}$ is the average tetrahedron edge length.

We define the local packing fraction $\Phi_{loc}$ and local contact number $Z$ as two possible order parameters for the free volume theory. We also calculate standard bond orientational order (BOO) parameters that have been commonly used to identify local crystalline or icosahedral orders[55]. Additionally, in our previous study, we found that there exists correlation between the local anisotropy index $\beta_1^{0,2}$ of the shape of the Voronoi cell and certain locally favored structures (LFSs) with five-fold symmetry[19]. We therefore also include it as a possible order parameter.

Local packing fraction $\Phi_{loc} = \dfrac{V_g}{V_{voro}}$.

Local contact number $Z$. Neighboring particles whose center-to-center distance to one particle is shorter than a threshold $r_c = 1.011\sigma$ are defined as its contacting neighbors. $\sigma$ is the average particle diameter. $r_c$ is determined from a previously-established analysis protocol.

Bond orientational order (BOO). We use a modified definition of BOO in which each bond is weighed by the area of its corresponding Voronoi facet[56]: $Q_{lm} = \sum_{i=1}^{k} \dfrac{A_i}{A} Y_{lm}(\theta_i, \varphi_i)$ where $(\theta_i, \varphi_i)$ is the angular position of the $i^{th}$ Voronoi neighbor in the spherical system of the central particle, $A_i$ is the area of the Voronoi facet shared by the central particle and its $i^{th}$ neighbor, and $A$ is the total surface of the Voronoi cell. $Y_{lm}$ are spherical harmonics. The local BOO parameters are defined as: $q_l = \sqrt{\dfrac{4\pi}{2l+1} \sum_{m=-l}^{l} |Q_{lm}|^2}$ and $w_l = \sum_{\substack{m_1,m_2,m_3 \\ m_1+m_2+m_3=0}} \begin{pmatrix} l & l & l \\ m_1 & m_2 & m_3 \end{pmatrix} Q_{lm_1} Q_{lm_2} Q_{lm_3}$, where $\begin{pmatrix} l & l & l \\ m_1 & m_2 & m_3 \end{pmatrix}$ are Wigner 3j symbols. Specifically, we calculated the BOO parameters $q_4$, $q_6$, $\hat{w}_4$ and $\hat{w}_6$, where $\hat{w}_l = w_l \Big/ \left( \sum_{m=-l}^{l} |Q_{lm}|^2 \right)^{3/2}$.

The local Voronoi cell anisotropy index $\beta_1^{0,2}$. It is calculated using a Minkowski tensor $W_1^{0,2} = \int \mathbf{n} \otimes \mathbf{n} \, dA$, defined as the surface integral of the tensor-valued self-product of the bounding surface normal $\mathbf{n}$ (ref. 57). $\beta_1^{0,2}$ is defined as the ratio of the smallest and largest eigenvalues of $W_1^{0,2}$. The value of $\beta_1^{0,2}$ ranges from one (isotropic shape) to zero (a line or a plane).

The purpose of the percolation analyses is to identify the spatial correlation properties of these local structural order parameters, that is, whether they form clusters or distribute randomly. To achieve this goal, we color cells based on the values of above structural order parameters. Although the exact nature of the amorphous structural order remains unknown, it is reasonable to assume they are locally compact. Therefore, they have smaller $\delta$ and $\Delta$, and larger $\Phi_{loc}$ and $Z$. For other structural order parameters, we calculated their Pearson correlation coefficients with $\Phi_{loc}$ and found that $\beta_1^{0,2}$ and $q_6$ show positive correlations, whereas $q_4$, $\hat{w}_4$ and $\hat{w}_6$ show negative correlations. Therefore, larger $\beta_1^{0,2}$ and $q_6$, and smaller $q_4$, $\hat{w}_4$ and $\hat{w}_6$ correspond to more compact local structures.

The coloring process works as follows: cells or simplexes with $x<(>)x_c$ (< for $\delta$, $\Delta$, $q_4$, $\hat{w}_4$ and $\hat{w}_6$ while > for $\Phi_{loc}$, $Z$, $\beta_1^{0,2}$ and $q_6$) are colored, where $x = \delta, \Delta, \Phi_{loc}, \beta_1^{0,2}, q_4, q_6, \hat{w}_4, \hat{w}_6$ is one of the above structural order parameters, and $x_c$ is the corresponding threshold. To compare with a random-coloring process, for each $x_c$ value, we also color the same number of cells or simplexes as those with $x<(>)x_c$, except that these cells are chosen randomly.

**Acknowledgements** We thank Walter Kob, Daan Frenkel and Hernán Makse for useful discussions, and Walter Kob also for critical review of the manuscript. Experiments were carried out at BL13W1 beamline of the Shanghai Synchrotron Radiation Facility and 2BM beamline of the Advanced Photon Source at Argonne National Laboratory. The use of the Advanced Photon Source is supported by the US Department of Energy, Office of Science, Office of Basic Energy Sciences, under Contract No. DE-AC02-06CH11357. The work is supported by the National Natural Science Foundation of China (No. 11175121 and U1432111) and National Basic Research Program of China (973 Program; 2010CB834301).

**Author Contributions** Y.W. designed research. C.X., J.L., Y.C., B.K., X.X., K.F., T.X. and Y.W. performed the experiment. C.X., J.L., Y.C., B.K. and Y.W. analyzed the data and all authors wrote the paper.

**Author Information** The authors declare no competing financial interests. Correspondence and requests for materials should be addressed to Y.W. (yujiewang@sjtu.edu.cn).

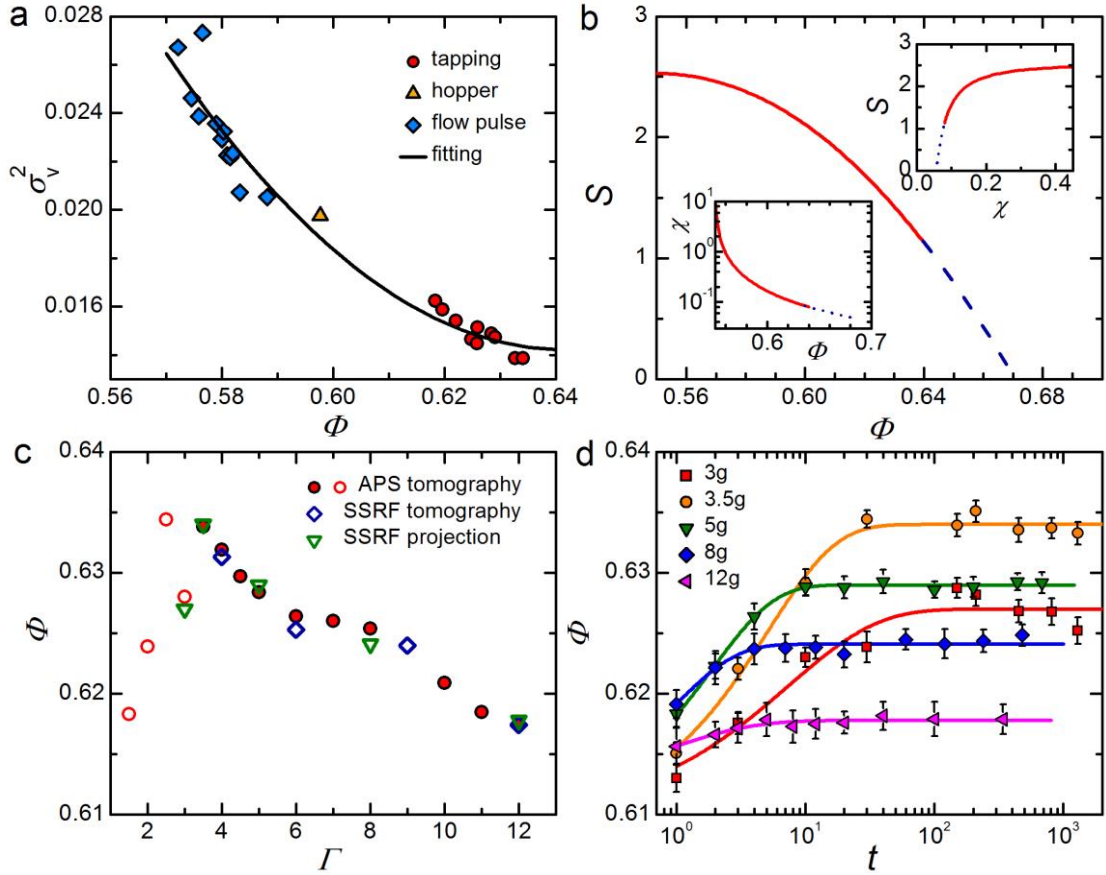

**Figure 1 | Dynamics and thermodynamics of tapped granular packing. a,** Variance of reduced Voronoi volume versus average packing fraction $\Phi$. Different symbols represent different packing preparation protocols: tapping (circles), hopper (triangles) and flow pulse (diamonds). The solid line is a second-order polynomial fitting. **b,** Entropy $S$ versus $\Phi$. The insets shows $S$ versus compactivity $\chi$ and the equation of state $\chi(\Phi)$. The dashed lines correspond to extrapolations towards $S=0$. **c,** Packing fraction $\Phi$ measured by tomography (circles and diamonds) and projection imaging (triangles) versus $\Gamma$. Open circles mark the end of the reversible branch where the experimental tapping number is insufficient for the system to reach steady states. **d,** Compaction curves of various tapping intensities $\Gamma$ and KWW fits (lines). Error bars are defined as s.e.m.. KWW, Kohlrausch-Williams-Watts.

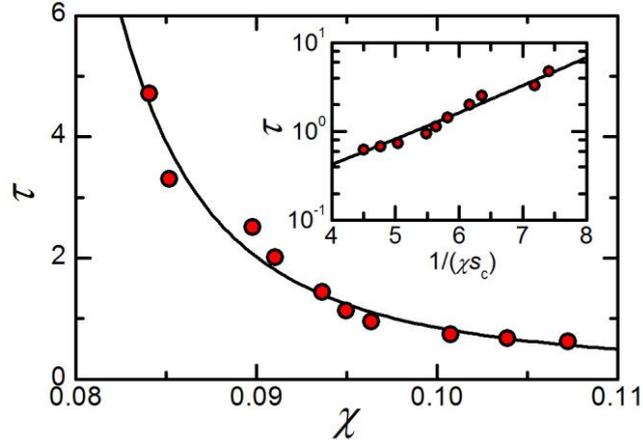

**Figure 2 | Relationships between relaxation time and thermodynamic variables.** Relaxation time $\tau$ versus $\chi$, and VFT fitting according to equation (3) (line). In this fitting, we fix the fragility index $D=4$ by adopting the hard-sphere simulation results (note the equivalence of $\chi$ and $T/P$ )[25,54], and obtain: $\tau_0 = 0.024 \pm 0.014$ and $\chi_0 = 0.049 \pm 0.003$. The inset shows $\tau$ versus $1/(\chi s_c)$ and fitting according to equation (4) (with $\tau_0 = 0.024$) (line). VFT, Vogel-Fulcher-Tammann.

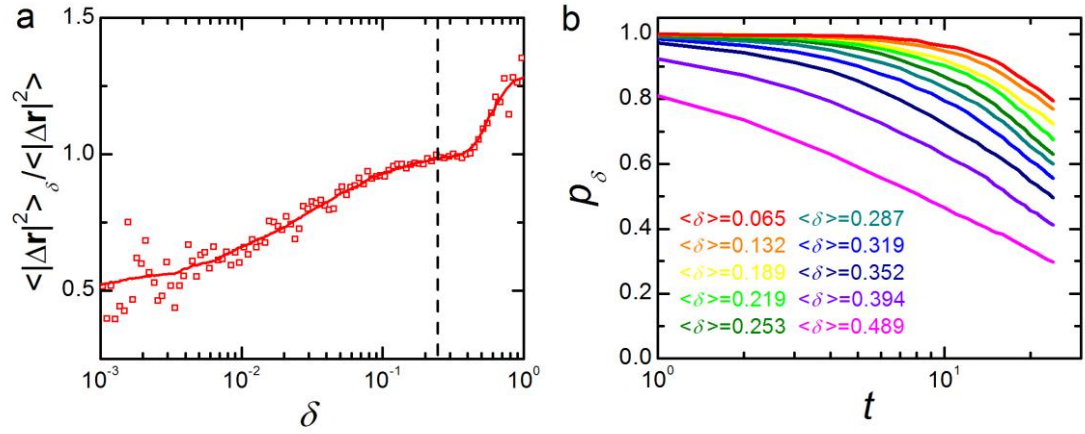

**Figure 3 | Correlations between slow dynamics and the polytetrahedral order. a,** normalized $\delta$–mobility $\Delta r^2$ versus $\delta$. Tetrahedra are classified into several groups based on their $\delta$ values to calculate their average mobility. The solid line is a guide to eye, and the dashed line marks the position of the threshold $\delta^*$ in defining quasi-regular tetrahedra. **b,** Tetrahedron correlation function $p_\delta(t)$. Lower curves correspond to tetrahedra with increasingly larger $\delta$ value at $t=0$.

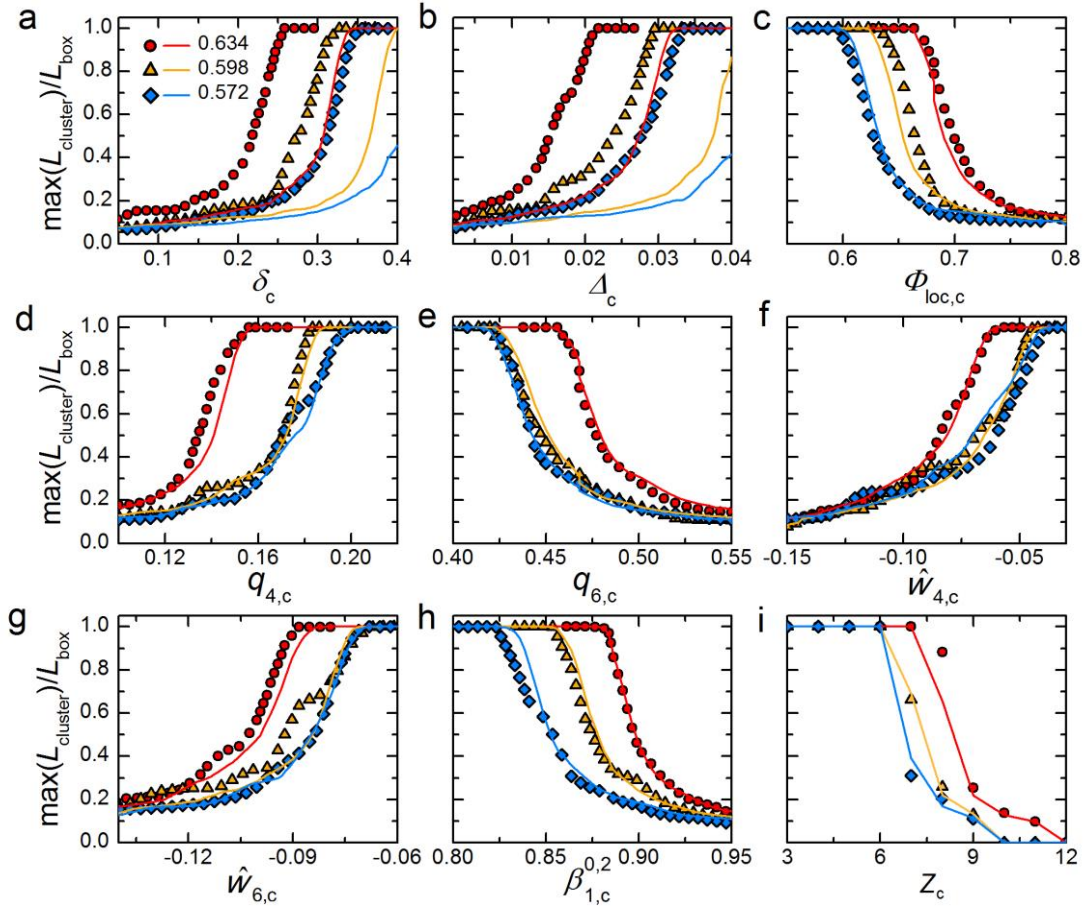

**Figure 4 | Percolation analyses of various structural orders.** Ratios between the maximum spanning range of clusters and box size as a function of the threshold of various structural order parameters (symbols) or random percolation cases (solid lines). (**a**, **b**) Polytetrahedral order. (**c**) Local packing fraction. (**d**-**g**) Bond orientational orders. (**h**) Local Voronoi cell anisotropy index. (**i**) Local contact number. Data for three different $\Phi$: $\Phi = 0.634$ (circles), $0.598$ (triangles) and $0.572$ (diamonds) are shown.

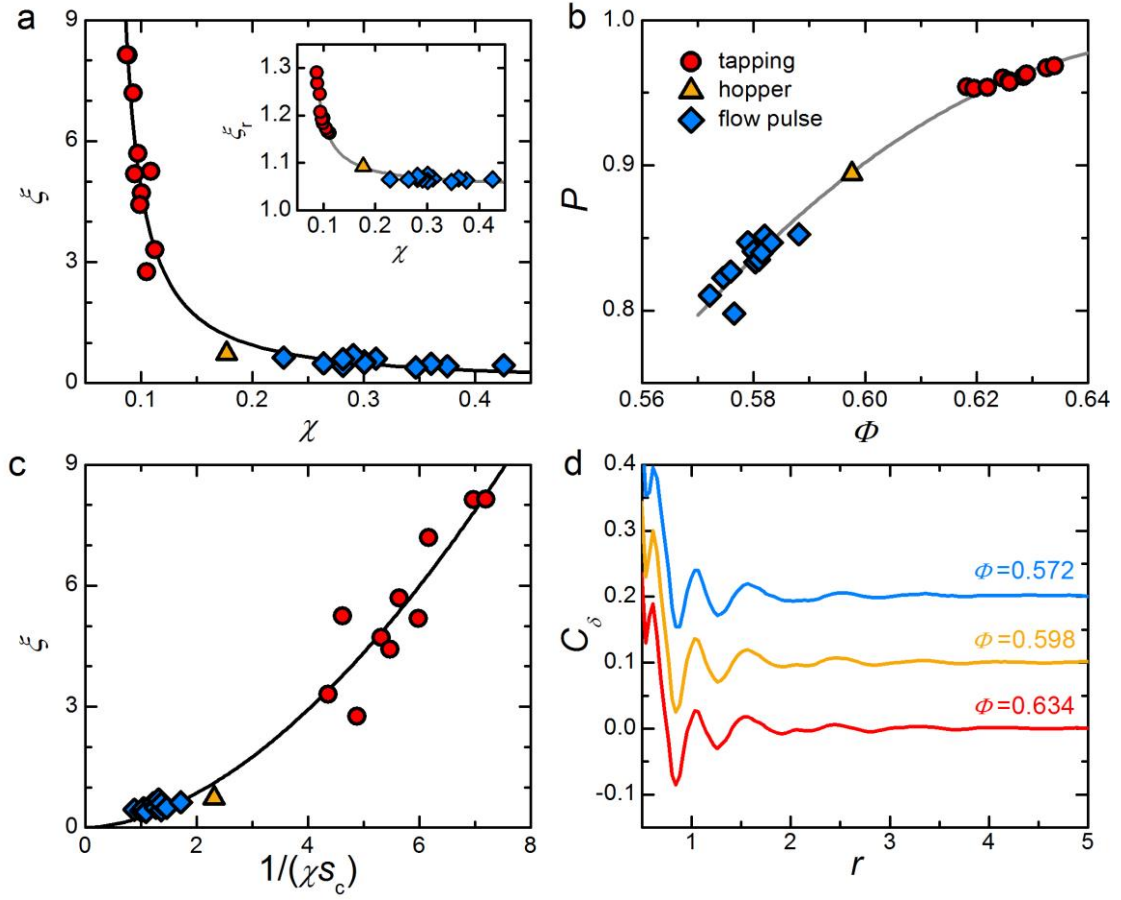

**Figure 5 | Correlation length of the polytetrahedral order. a,** Correlation length $\xi$ versus compactivity $\chi$ and fitting according to equation (5) (line). The inset shows $\xi_r$ versus $\chi$, and the solid line is a guide to eye. **b,** $P$ as a function of $\Phi$. The solid line is a guide to eye. **c,** $\xi$ versus $1/(\chi s_c)$ and fitting according to equation (6) (line). In **a, b, c,** the meanings of the symbols are the same as those in Fig. 1a. **d,** Spatial correlation functions of

$$\delta: \; c_\delta(r) = \left\langle \frac{(\delta_i - \langle \delta \rangle)(\delta_j - \langle \delta \rangle)}{\mathrm{var}(\delta)} \right\rangle,$$ where the average is taken for each pair of tetrahedra separated by distance $r$, and $\delta_i$ and $\delta_j$ are their $\delta$ values. The location of a tetrahedron is defined as the average location of its four particles. The lines are shifted for clarity.

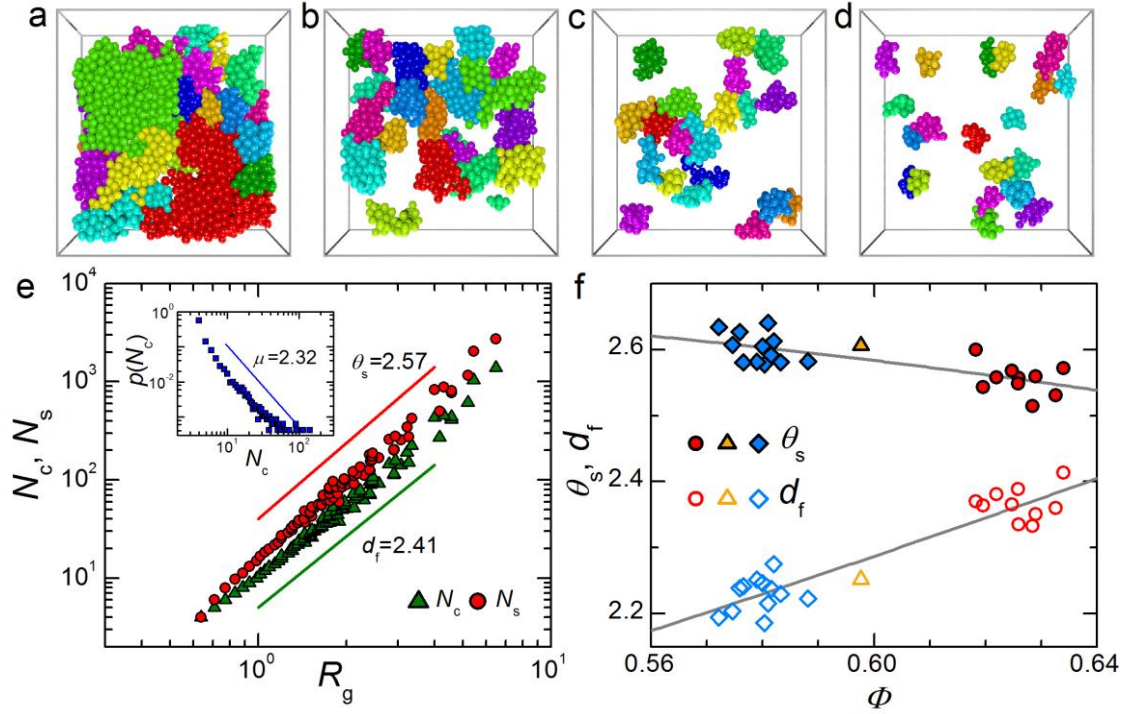

**Figure 6 | Structure of the polytetrahedral order. a, b, c, d,** Configurations of the largest twenty polytetrahedral clusters for packing of $\Phi = 0.634$ (**a**), $0.618$ (**b**), $0.598$ (**c**), and $0.572$ (**d**). **e,** The surface area $N_s$ (circles) and cluster size $N_c$ (squares) versus radius of gyration $R_g$. $N_s$ is defined as the number of tetrahedra face-adjacent to a polytetrahedral cluster formed by quasi-regular tetrahedra. The inset shows the probability of finding a cluster with size $N_c$. Only data for $\Phi = 0.634$ is shown. The solid lines mark the slopes of corresponding scaling behaviors: $N_s \propto R_g^{\theta_s}$, $N_c \propto R_g^{d_f}$ and $p(N_c) \propto N_c^{-\mu}$. **f,** $\theta_s$ and $d_f$ versus $\Phi$. The solid lines are guides to eye. The meanings of the symbols are the same as those in Fig. 1a.